\documentclass[twocolumn,apl,preprintnumbers,amsmath,amssymb,altaffillsymbol]{revtex4}
\usepackage{graphicx}% Include figure files
\usepackage{dcolumn}% Align table columns on decimal point
\usepackage{bm}% bold math
\usepackage{subfigure}
\usepackage{color}
\usepackage{amsmath}

\begin{document}

\title{Dynamic Interactions between Oscillating Cantilevers: \\ Nanomechanical Modulation using Surface Forces}
\author{O. Basarir}
\author{K. L. Ekinci}
\altaffiliation{ Author to whom correspondence should be addressed; electronic mail: ekinci@bu.edu}
\affiliation{ Mechanical Engineering Department and the Photonics Center, Boston University, Boston, Massachusetts 02215}

%_______________________________________________________________________________________________________________________
% ABSTRACT

\begin{abstract}

Dynamic interactions between two oscillating micromechanical cantilevers are studied. In the experiment, the tip of a high-frequency cantilever is positioned near the surface of a second low-frequency cantilever. Due to the highly nonlinear interaction forces between the two surfaces, thermal oscillations of the low-frequency cantilever modulate the driven oscillations of the high-frequency cantilever. The dissipations and the frequencies of the two cantilevers are shown to be coupled, and a simple model for the interactions is presented. The interactions studied here may be useful for the design of future micro and nanoelectromechanical systems for mechanical signal processing; they may also help realize coupled mechanical modes for experiments in non-linear dynamics.

\end{abstract}

%_______________________________________________________________________________________________________________________
% INTRODUCTION

\maketitle

Miniaturized mechanical devices \cite{Ekinci}, called micro and nanoelectromechanical systems (MEMS and NEMS), are steadily progressing toward attaining the speed and efficiency of their electronic counterparts. A mechanical signal processor \cite{blick} may soon become a reality if the primary operations during the processing can be performed mechanically --- \emph{i.e.}, using the mechanical motion of MEMS and NEMS. There is thus a focused research effort to realize micro- and nano-mechanical  switching \cite{ho-bun}, mixing \cite{roukes_piezo}\cite{ron_lifshitz}\cite{capacitive_ieee}, amplification \cite{inna} and modulation\cite{rf_stm,hiebert-freeman}.

Here, we study dynamic interactions between two oscillating micromechanical cantilevers and harvest these interactions for mechanical signal modulation and detection. In the experiment, the \emph{carrier signal} from a high-frequency  microcantilever oscillator is modulated by  low-frequency thermal oscillations of a second microcantilever by simply bringing the two  microcantilevers close together. The approach relies upon the strong inherent nonlinearity of the interaction force between two surfaces in close proximity and offers several advantages. The modulation is purely mechanical, and mechanical signals need not be converted to electrical signals.  The strength of the coupling between the two mechanical signals, and hence the modulation index, can be adjusted by changing the distance between the two microcantilevers. Conservative and dissipative components of the interaction  enable tuning of the frequencies and dissipation. Conversely, monitoring the modulation on the carrier signal allows for sensitive mechanical displacement detection. Because the approach offers prospects for creating coupled mechanical modes \cite{paul-ducker} with tunable coupling, it may be useful in fundamental investigations in nonlinear dynamics.

Our approach is derived from dynamic mode atomic force microscopy (AFM). Related to our work here, various AFM modalities have been used to detect the motion of micro- and nano-mechanical resonators. In these experiments, a resonant mode of the small device under study is excited, \emph{e.g.}, electrostatically or by using a piezoelectric shaker. An AFM cantilever is brought in close proximity of the resonator to probe its oscillations. Several groups have used contact interactions \cite{safar_resonator,Sensors_AFM,paulo_static}; tapping mode AFM has also been employed for less intrusive probing \cite{paulo_dynamic, garcia_prl_nanotubes,Craighead_afm}. In addition, other AFM-based approaches, including acoustic force microscopy \cite{SAFM_Sthal} and electrostatic scanning probe microscopy \cite{JMEMS_parametric}, have also been applied to measuring small oscillations. In this work, we extend the above-mentioned  efforts to non-contact mode AFM and show that even non-contact AFM can perturb small resonators significantly.

%_______________________________________________________________________________________________________________________
% EXPERIMENTAL SYSTEM

\begin{figure}[t]
\centering
\includegraphics[width=3.3 in]{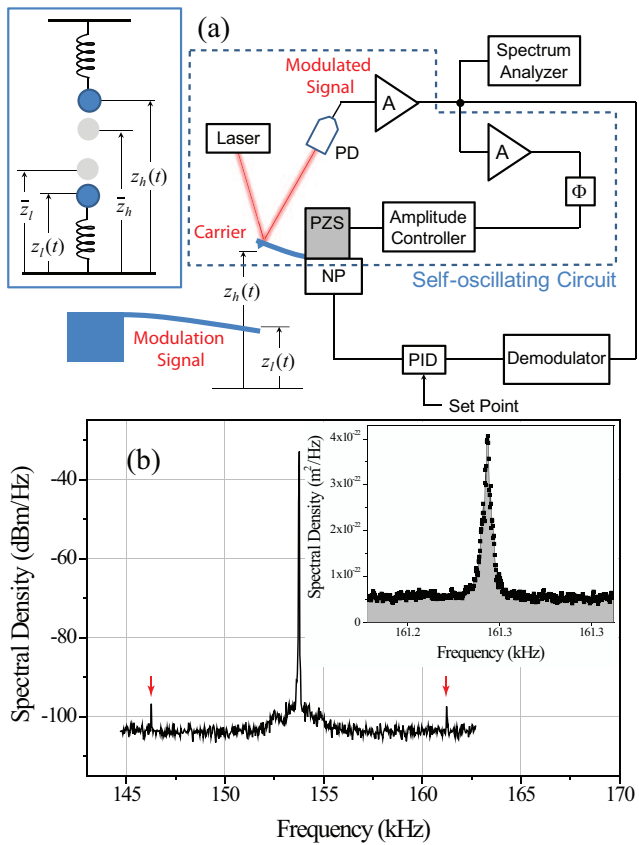}
\caption{\textbf{(a)} Schematic of the experimental setup. The optical transducer (PD: photodetector), amplifiers (A),  phase shifter ($\Phi$) and amplitude controller are the components of a  positive feedback circuit (dashed box), which drives the high-frequency cantilever at resonance at a prescribed amplitude via a piezo-shaker (PZS). The demodulator and PID controller form a  detection circuit, which keeps the frequency shift (and hence the average gap between cantilevers) at a desired value. The inset shows the lumped mass models for the two cantilevers. \textbf{(b)} Spectral density of the high-frequency cantilever oscillations. Two arrows at the upper and lower sidebands of the carrier at 153.8 kHz correspond to the thermal oscillations of the low-frequency cantilever.  The inset shows the upper sideband in displacement units. }
\label{fig:set-up}
\end{figure}

At the core of our experiment are two micro-cantilevers, which primarily interact through non-contact forces, as shown in Fig.~\ref{fig:set-up}(a). The cantilevers  are of different sizes and oscillate at their fundamental flexural resonance frequencies. The  smaller high-frequency one  (hereafter labeled with the subscript ``h'') has an \emph{unperturbed} fundamental flexural resonance at $f_{h0} \approx153.8$ kHz. The larger low-frequency cantilever (hereafter labeled with the subscript ``l'') comes with unperturbed $f_{l0}\approx10.1$ kHz. The \emph{in vacuo} parameters for both  cantilevers are listed in Table I.  The cantilevers  remain inside an ultrahigh vacuum (UHV) chamber at a pressure $p<7\times10^{-10}$ Torr during the experiments so that gas damping is not relevant \cite{ekinciLoC10}. The low-frequency (bottom) cantilever is fixed onto a sample holder and is excited by thermal fluctuations at room temperature. The  high-frequency (top) cantilever sits   on  a nano-positioner and  is driven  by a piezo-shaker at its base. The response of the high-frequency cantilever is measured using a standard optical beam-deflection method \cite{Meyer_Optic}. The output of the optical transducer is divided between a spectrum analyzer, a self-oscillating loop and a  detection-feedback feedback loop, as shown in Fig.~\ref{fig:set-up}(a). The self-oscillating loop, shown by the dashed box in Fig.~\ref{fig:set-up}(a), maintains the high-frequency cantilever oscillating at resonance at a constant amplitude.  The detection-feedback loop has a large time constant (0.01 s $\lesssim\tau\lesssim$ 1 s)  as compared to  other time scales in the experiment. It thus keeps the average gap between the cantilevers at a prescribed value and compensates for drifts.
%_______________________________________________________________________________________________________________________
% EXPERIMENTAL SETUP FIGURE

In the experiments, the high-frequency cantilever oscillates coherently at a constant r.m.s oscillation amplitude  of $\sim 10$ nm; the  r.m.s. thermal  oscillation amplitude for the  low-frequency cantilever remains around $\sqrt{k_BT \over k_l}\sim 0.1$ nm, where $k_BT$ is the thermal energy and $k_l$ is the (unperturbed) spring constant. The tip of the high-frequency cantilever is  brought towards the free end of the low-frequency cantilever using the nano-positioner, and the spectrum of the oscillatory signal on the photodiode is measured. Fig.~\ref{fig:set-up}(b) shows a typical spectral density measurement. The dominant peak shown at  $f_h\approx153$ kHz corresponds to the self-oscillations of the high-frequency cantilever. This can be regarded as the high-frequency carrier signal. At $f_h\pm f_l\approx153\pm 8$ kHz, two small peaks are noticeable. These upper and lower sideband modulation peaks result from the  thermal oscillations of the low-frequency cantilever. The  inset shows a close-up of the upper sideband peak in linear scale. Under these experimental conditions, using the thermal oscillation amplitude, we  calculate the noise floor for displacement detection  to be  $ \sim 1 \times 10^{-11}$ m/Hz$^{1/2}$.

\begin{table}[b]
\caption{\label{tab:devices} Unperturbed parameters for the two Silicon microcantilevers used in our experiments. The stiffness $k$ values are provided by the manufacturer. The effective mass $m$ is calculated using $k$ and $f_0$. Both $k$ and $m$ are approximate.}
\begin{ruledtabular}
\begin{tabular}{ccccccc}

$l \times w \times t$ & $f_{0}={{\omega_{0}}\over{2\pi}} $ & $Q_0$  &  $k$ & $m$ \\
 $\mu$m$^{3}$ & kHz &     & N/m  & kg \\
\hline
$225\times37.5\times7$ &  153.8 & $2 \times 10^4$ &    $30$ & $3\times10^{-11}$ \\
$450\times50\times2$ &  10.1 & $6\times 10^3$ &   $0.2$ & $5\times10^{-11}$ \\
\end{tabular}
\end{ruledtabular}
\end{table}

Advancing the nano-positioner leads to changes in the actual gap between the two cantilevers, resulting in changes in the measured response. Returning to the inset of Fig.~\ref{fig:set-up}(a), we note that the \emph{time-dependent} positions of the two cantilevers are   $z_h(t)$ and $z_l(t)$ with respect to a fixed reference point. The interaction force $F$, which has an attractive van der Waals component (see below for a details), results in changes in the \emph{average} positions, $\bar z_l$ and $\bar z_h$. In our coordinate system, the  average gap is $\approx \bar z_h - \bar z_l$. Because the low-frequency cantilever is two orders of magnitude softer than the high-frequency cantilever, we estimate that the average van der Waals attraction mostly bends the low-frequency cantilever toward the high-frequency cantilever (upwards in Fig.~1(a)) as the nano-positioner is advanced to decrease the gap between the cantilevers (i.e., to reduce $\bar z_h$). The average position of the high-frequency cantilever, $\bar z_h$, can  be taken to be the same as that of the nano-positioner (up to an additive constant).

\begin{figure*}
\centering
\includegraphics[width=7in]{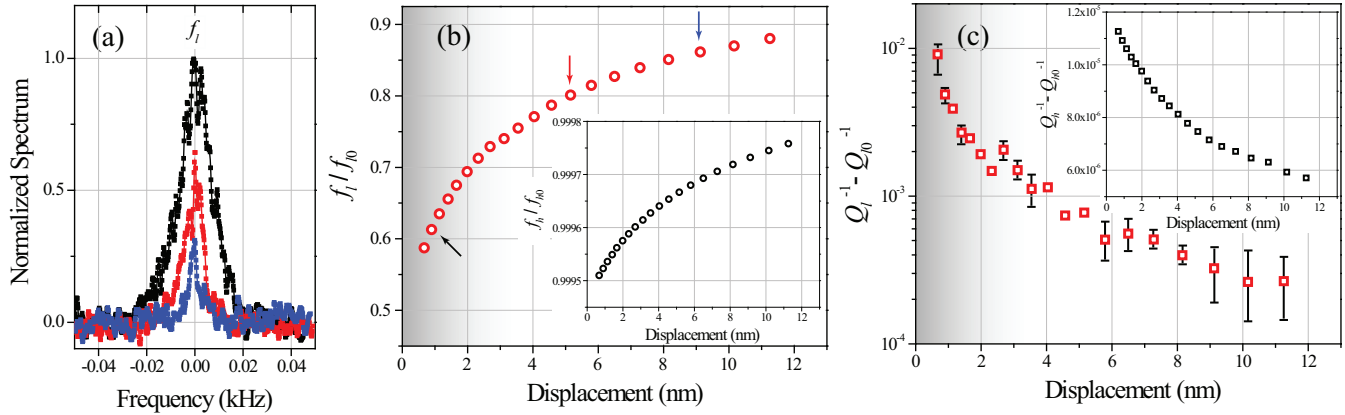}
\caption{\textbf{(a)} Normalized sideband signals. The signals are normalized using the highest measured signal values. The data traces are taken at the positions shown with the arrows in \textbf{(b)}. Because the resonance frequency $f_l$ shifts significantly, the frequency axis is displayed as measured from the resonance frequency $f_l$. \textbf{(b)} The observed shift in the resonance frequencies of both cantilevers. \textbf{(c)} The change in the dimensionless dissipation of both cantilevers. The dissipation increases dramatically in the shaded region,  suggesting that soft contact interactions start to become dominant. The error bars in all the data are smaller than the symbol sizes unless shown explicitly. The snap to contact with accompanying instabilities in the high-frequency signal determines the position of zero in the $x$-axes of the plots in (b) and (c).}
\label{fig:pressure-topographs}
\end{figure*}

When far away from each other, the cantilevers do not interact and oscillate at their respective unperturbed resonance frequencies, $f_{h0}$ and $f_{l0}$. As they come closer, the coupling grows and the thermal oscillations of the low-frequency cantilever become observable in the sidebands of the carrier. As the separation is reduced, the linewidths and the frequencies of both cantilevers change.  The motion of the high-frequency cantilever (carrier) remains mostly sinusoidal with relatively little perturbation to its resonance frequency and linewidth, since the modulating signal in the sideband is orders of magnitude smaller than the carrier. The low-frequency cantilever, on the other hand, suffers large changes in frequency and linewidth. Fig. 2(a) shows the sideband peaks corresponding to the low-frequency cantilever oscillations. The zero in the frequency axis here corresponds to the resonance frequency $f_l$, which decreases as the nano-positioner advances  to bring the two cantilevers together. The modulation increases because the mechanical coupling increases. In addition, the dissipation (linewidth) increases.

Figure 2(b) and (c) show results   from systematic experiments as the nano-positioner brings the two cantilevers together, i.e.,  the gap between the cantilevers is changed. Returning to Fig. 1(a), we describe how the experiment is performed. A frequency shift for the high-frequency cantilever is prescribed; the nano-positioner (in conjunction with the detection circuit)  brings the two cantilevers together until this frequency set point is attained. The PID controller keeps this frequency shift (set point) fixed, thereby ensuring a constant average gap. At this set point, line-shape for the low-frequency cantilever is recorded. At very small separations, the two cantilevers  snap to hard contact, causing the carrier signal to become unstable.

In  Fig. 2(b) (main), the  frequency $f_l$ of the low-frequency cantilever is plotted as a function of the nano-positioner displacement. The inset of Fig. 2(b) similarly shows  $f_h$ of the high-frequency cantilever as a function of the nano-positioner displacement. The zeros of the $x$-axes are taken to be the contact position, where the high-frequency cantilever can no longer oscillate stably. There is some degree of uncertainty in the position of this zero. The estimated region of soft contact between the two cantilevers is shown by the shading around zero in the main figure. This estimation is simply based on the observation that the dissipation of both cantilevers increases more steeply for displacements $\lesssim 4$ nm. The data traces in Fig. 2(b) (main and inset)  showing negative frequency shifts (for both cantilevers) appear  qualitatively similar to the frequency shift \emph{vs.}  tip-sample distance curves taken in non-contact AFM work \cite{AFM_book}, where attractive forces are dominant. However, there is a significant difference. Because the low-frequency cantilever is soft,  the nominal displacement obtained from the nano-positioner cannot be related  to the  tip-sample gap in a straightforward manner. The interaction range in Fig.~2(b)  extends over 10 nm. Due to the attractive force between the two cantilevers, the soft cantilever follows the stiffer high-frequency cantilever, as the two are brought together.   While both resonance frequencies $f_l$ and $f_h$ shift in a qualitatively  similar fashion,  the magnitudes of the changes in $f_l$ and $f_h$ are quite different. We confirm that the data possess the same features at  larger oscillation amplitudes $\lesssim 40$ nm (not shown) of the high-frequency cantilever. In all the measurements reported here, non-contact or (intermittent) soft contact interactions dominate, and the average force between the cantilevers remains attractive.

Figure 2(c) shows the change in the dimensionless dissipation of each cantilever as a function of the nano-positioner displacement. Here, the change is obtained by subtracting the intrinsic value of the dimensionless dissipation, ${Q_0}^{-1}$, from the measured value $Q^{-1}$ for each cantilever. For the low-frequency cantilever, all the data points are obtained by fitting Lorentzians to resonance line-shapes, such as those shown in Fig. 2(a). At large separations between cantilevers, the data appears noisier. This is because the signal size becomes smaller, and the fits are not as accurate. For the high-frequency cantilever, the data are extracted from the drive force (voltage) applied to the piezo-shaker, given that the stiffness of the high-frequency cantilever does not change appreciably and the amplitude controller keeps the oscillation amplitude constant\cite{AFM_book}. The general trend is that dissipation increases as the separation decreases. However, the observed dissipation increase in the low-frequency cantilever is much more dominant.

%_______________________________________________________________________________________________________________________
% EQUATIONS OF MOTION

We now describe the coupled resonator dynamics. The dynamic variables used in the equations below can be identified in Fig. 1(a). Before  analyzing the interacting cantilevers, we formulate the dynamics of  individual cantilevers far apart from each other. The one-dimensional lumped equation of motion for the high-frequency cantilever can be written as ${m_h}{\ddot z_h} + {m_h}{{{\omega _{h0}}} \over {{Q_{h0}}}}{\dot z_h} + {m_h}{\omega _{h0}}^2\left({z_h-\bar z_h}\right) = {F_d}(t)$, where the drive force is $F_d(t)=R\left( z_h(t-t_{\phi})-\bar z_h \right)$, with $R$ being the gain, $t_{\phi}$ being the loop delay of the (self-oscillating) loop, and $m_h$ being the effective mass of the cantilever.  We use the simplifying assumption that the cantilever always vibrates sinusoidally at resonance at a constant amplitude, and the role of the external sustaining circuit is to simply   compensate for the energy losses. Thus, we can write $z_h(t)\approx \bar z_h + A_h \sin {\omega_{h0}}t$, where $A_h$ remains constant and $\omega_{h0}$ does not change appreciably, consistent with experimental observations [Fig. 2(b) inset]. The low-frequency cantilever is driven by random thermal noise, but oscillates mostly sinusoidally because of its high quality factor ($10^2 \le Q_l \lesssim 10^4$). Modeling its displacement as narrowband noise \cite{phase_noise}, we write  $z_l(t)\approx \bar z_l+ A_{l}(t) \sin (\omega_{l0} t+\psi_{l}(t))$, where $A_l(t)$ and $\psi_l(t)$ are  \emph{slowly-varying} envelope and phase functions. Hence, both cantilevers can be treated as simple one-dimensional oscillators when no perturbations are present: ${m_i}{\ddot z_i} + {m_i}{\omega _{i0}}^2\left( {{z_i} - {{\bar z}_i}} \right) \approx 0$, where $i=l,h$. Thus, both cantilevers oscillate sinusoidally  with $\omega _{h0} \gg \omega _{l0} $, and each will tend to respond strongly to the perturbation at its own resonance frequency. For our system, when the gap between the cantilevers is large, the generalized non-contact interaction force can be expressed in terms of the coordinates and their time derivatives:  $F=F( z_h, \dot z_h, z_l, \dot z_l)$ \cite{AFM_book}. This force can further be broken down into  conservative and dissipative components as $F=F_{\rm{diss}}+F_{\rm{cons}}$ \cite{cons_dissi_tip-sample}.

The dissipative forces ${F_{{\text{diss}}}}$ on the high-frequency  and low-frequency cantilevers can be approximated as \cite{cons_dissi_tip-sample}  $ -\gamma ({\dot z_h} - {\dot z_l})$ and $ -\gamma ({\dot z_l} - {\dot z_h})$, respectively,  based upon phenomenological  arguments. Here,  $\gamma$ is a function of gap: $\gamma({z_h},{z_l})= {\gamma _0}{e^{ - C({z_h} - {z_l})}}$, where $\gamma_0$ and $C$ are empirical constants. The exponentially decaying form ensures that $F_{\rm{diss}}$ becomes weaker with increasing separation. Interacting only via the dissipative force $F_{\rm{diss}}$, the two cantilevers can be described by the following coupled equations:
\begin{subequations}
    \begin{align}
        m_l\ddot z_l  + m_l{\omega _{l0}} ^2 \left( {{z_l} - {{\bar z_l}}} \right) \approx & -{\gamma}({\dot z_l} - {\dot z_h}),\\
        m_h\ddot z_h + {m_h\omega _{h0}}^2 \left( {{z_h} - {{\bar z}_h}} \right) \approx & -{\gamma}({\dot z_h} - {\dot z_l}).
    \end{align}
\label{eq:coupled-diss}
\end{subequations}
\noindent To make further progress, we approximate the function $\gamma(z_h,z_l)$ with $\bar \gamma  \approx {\gamma _0}{e^{ - C({\bar z_h} - {\bar z_l})}}$.  Because of the discrepancy in the two oscillatory time scales, the low frequency cantilever notices only the average position of the high-frequency cantilever, $z_h=\bar z_h$. It may thus be justifiable to set $\dot z_h\approx 0$ in Eq.~\ref{eq:coupled-diss}(a). This results in $m_l\ddot z_l  + \bar \gamma {\dot z_l} + m_l{\omega _{l0}} ^2 \left( {{z_l} - {{\bar z}_l}} \right) \approx 0$. Similarly, the dissipative force acting on the high-frequency cantilever is approximately  $-{\bar \gamma}{\dot z_h}$ because $A_h \omega _{h0} \gg A_l \omega _{l0} $. Thus, we arrive at the approximation $m_h\ddot z_h + \bar \gamma {\dot z_h} + {m_h\omega _{h0}}^2 \left( {{z_h} - {{\bar z}_h}} \right) \approx 0$. It can be seen that $\bar \gamma$ terms give rise to the energy dissipation in both cantilevers. Thus,  one should be able to relate the measured dissipation changes in the coupled cantilever system. In other words, ${m_h}{\omega _h}\left( {{Q_h}^{ - 1} - {Q_{h0}}^{ - 1}} \right)\sim {m_l}{\omega _l}\left( {{Q_l}^{ - 1} - {Q_{l0}}^{ - 1}} \right)$ at a given gap. At the largest gap values, where the perturbation is weak and the approximations should hold better, we find the right hand side and the left hand side to be of the same order of magnitude ($\rm{r.h.s}\approx 2\times 10^{-10}$ kg/s and $\rm{l.h.s}\approx6\times10^{-10}$ kg/s) using the numbers from Table 1 and data from Fig. 2(c). Given that the values in Table 1 are approximate, this is quite satisfactory and suggests that our approximations are reasonable.

Returning now to the conservative component of the interaction force, we take $F_{\rm{cons}} = -\frac{{HR}}{{{{\left( {{z_h} - {z_l}} \right)}^2}}}$, as suggested by numerous AFM experiments \cite{cons_dissi_tip-sample}. Here, $H$ is the Hamaker constant, and $R$ is the tip radius (of the high-frequency cantilever). We emphasize that this simple form is valid when the gap is large (non-contact regime), and the attractive  van der Waals force dominates. Because the thermal  oscillation amplitude (of the low-frequency cantilever) remains extremely small, we expand the force around  $\bar z_l$, obtaining
\begin{equation}
{\tilde F_{{\text{cons}}}} \approx  - \frac{{HR}} {{{{\left( {{z_h} - {{\bar z}_l}} \right)}^2}}} - \frac{{2HR}} {{{{\left( {{z_h} - {{\bar z}_l}} \right)}^3}}}\left( {{z_l} - {{\bar z}_l}} \right).
\label{eq:expanded-cons-force}
\end{equation}
Note that the sign of $\tilde F_{{\text{cons}}}$ must be adjusted such that it remains attractive for both cantilevers. As above, we set $z_h=\bar z_h$ in ${\tilde F_{{\text{cons}}}}$ in the equation of motion of the low-frequency cantilever: $m_l\ddot z_l + {m_l\omega _{l0}}^2 \left(z_l- \bar{z_l}\right) \approx {\tilde F_{{\text{cons}}}}$. This yields
\begin{equation}
{\omega_{l}}^2 \approx {\omega _{l0}}^2 - {{2HR} \over {m_l {{\left( {{{\bar z}_h} - {{\bar z}_l}} \right)}^3}}}.
\label{eq:freq-shift-low}
\end{equation}
Finally, the source of the modulation can be identified as the ${{2HR\left( {{z_l} - {{\bar z}_l}} \right)} \over {{{\left( {{z_h} - {{\bar z}_l}} \right)}^3}}}$ term in the drive force in $m_h\ddot z_h + {m_h\omega _{h0}}^2 \left( z_h-\bar z_h \right) \approx -{\tilde F_{{\text{cons}}}}$.  This term can be re-written as ${{2HR({z_l} - {{\bar z}_l})} \over {{{({{\bar z}_h} - {{\bar z}_l})}^3}{{\left( {1 + {{{z_h} - {{\bar z}_h}} \over {{{\bar z}_h} - {{\bar z}_l}}}} \right)}^3}}}$ and expanded, with the leading order term in  $\left( {{z_h} - {{\bar z}_h}} \right)$ being $- {{6HR\left( {{z_l} - {{\bar z}_l}} \right)\left( {{z_h} - {{\bar z}_h}} \right)} \over {{{\left( {{{\bar z}_h} - {{\bar z}_l}} \right)}^4}}}$. Including this term in the equation of motion, we derive
\begin{equation}
 {\omega _h}^2 \approx {\omega _{h0}}^2 - {{6HR} \over {{m_h}{{({{\bar z}_h} - {{\bar z}_l})}^4}}}{A_l}\sin ({\omega _{l0}}t + {\psi _l}).
\label{eq:freq-shift-high}
\end{equation}
 This is the source of the frequency modulation. The modulation index can be found as \cite{phase_noise} $M={{3HR{A_l}} \over {{m_h}{\omega _{h0}}{\omega _{l0}}{{({{\bar z}_h} - {{\bar z}_l})}^4}}}$, with the ratio of the power in the carrier to that in a (single) sideband being $M^2\over 4$. For the measurement shown in Fig.~\ref{fig:set-up}(b),  ${M^2\over 4}\approx 1.8\times10^{-3}$. Using $H\sim10^{-19}$ J, $R\sim 50$ nm and experimental values for the remaining parameters, we find ${\bar z_h} - {\bar z_l}\sim1$ nm.

More experimental and theoretical work is needed for a better understanding of this interesting coupled system. From an experimental perspective, a direct measurement of the gap between the cantilevers may be important. In the model, we  assume that the amplitude $A_h$ stays constant and the cantilevers oscillate sinusoidally. To fully account for the nonlinear interaction, a better model must allow for the amplitudes to be affected, with some degree of amplitude modulation as well as nonlinearity in the oscillations. Furthermore, the presented model is expected to   become inaccurate as the perturbation grows (\emph{i.e.}, the gap becomes smaller), and the  dynamics becomes complicated due to stronger non-linearities, hysteresis and larger fluctuations. One can incorporate contact effects by using  Derjaguin-M\"{u}ller-Toporov interaction.  Regardless, the data and the simple model presented here may be useful for designing MEMS and NEMS devices for future applications.  Given that the interaction  between the two cantilevers can be tuned efficiently by reducing the gap between them, one can also study non-linear dynamics of coupled oscillators.

We acknowledge support from the US NSF (through Grant Nos. ECCS-0643178 and  CMMI-0970071).

\end{document}